%% file: main-arxiv.tex
\documentclass[letterpaper,twocolumn,10pt]{article}
\usepackage{usenix}
\usepackage{fancyhdr}
\usepackage{tikz}
\usepackage{cite}
\usepackage{amsmath,amssymb,amsfonts}
\usepackage{algorithmic}
\usepackage{graphicx}
\usepackage{textcomp}
\usepackage{xcolor}
\usepackage{comment}
\usepackage[tightlists,tightblockquotes]{wspr}
\usepackage{fixedvalues}
\usepackage{booktabs}
\usepackage{multirow}
\usepackage{threeparttable}
\usepackage{hyperref}
\usepackage{url}
\usepackage{tablefootnote}
\usepackage[labelfont=bf,font=normal,aboveskip=2pt,belowskip=2pt]{caption}

\usepackage{titlesec}
\titlespacing{\section}{0pc}{8pt}{8pt}
\titlespacing{\subsection}{0pc}{6pt}{6pt}

\newenvironment{myquote}%
  {\list{}{\leftmargin=0.1in\rightmargin=0.1in}
  \vspace{-4pt}
  \item[]
  }%
  {\endlist
    \vspace{-4pt}
    }
\AtBeginEnvironment{myquote}{\itshape}

\newcommand{\paragraphb}[1]{\vspace{0.05in}\noindent{\bf #1} }

\begin{document}
\date{}

\title{\Large \textbf{ Enabling Developers, Protecting Users: Investigating Harassment and Safety in VR}
}

\author{
{\rm Abhinaya S.B.}\\
\rm North Carolina State University\\
asrivid@ncsu.edu
\and
{\rm Aafaq Sabir}\\
\rm North Carolina State University\\
asabir2@ncsu.edu
\and
{\rm Anupam Das}\\
\rm North Carolina State University\\
anupam.das@ncsu.edu
}

\maketitle

\input{sections/00-abstract}
\input{sections/01-intro}

\input{sections/02-relwork}

\input{sections/03-method}
\input{sections/04-results-01}
\input{sections/04-results-02}
\input{sections/05-results-03}
\input{sections/06-discussion}
\input{sections/07-conclusion}
\input{sections/08-ack}

\bibliographystyle{IEEETran-bib}
{\scriptsize 
\input{main-arxiv.bbl}

}
\input{sections/09-appendix}
\end{document}

%% file: sections/00-abstract.tex
\begin{abstract}
Virtual Reality (VR) has witnessed a rising issue of harassment, prompting the integration of safety controls like muting and blocking in VR applications. However, the lack of standardized safety measures across VR applications hinders their universal effectiveness, especially across contexts like socializing, gaming, and streaming. While prior research has studied safety controls in social VR applications, our user study (n = $\NumberOfInterviews$) takes a multi-perspective approach, examining both users' perceptions of safety control usability and effectiveness as well as the challenges that developers face in designing and deploying VR safety controls. We identify challenges VR users face while employing safety controls, such as finding users in crowded virtual spaces to block them. VR users also find controls ineffective in addressing harassment; for instance, they fail to eliminate the harassers' presence from the environment. Further, VR users find the current methods of submitting evidence for reports time-consuming and cumbersome. Improvements desired by users include live moderation and behavior tracking across VR apps; however, developers cite technological, financial, and legal obstacles to implementing such solutions, often due to a lack of awareness and high development costs. We emphasize the importance of establishing technical and legal guidelines to enhance user safety in virtual environments.
\end{abstract}

%% file: sections/01-intro.tex
\section{Introduction}
\label{sec:intro}

\noindent\framebox[\linewidth][c]{\begin{minipage}[t]{0.95\linewidth} \textbf{Content Warning:} This paper studies harassment in VR. This paper directly quotes participants when necessary, which may contain descriptions of offensive/hateful speech, profanity, and other potentially triggering content. \end{minipage}}
\vspace{2pt}

\noindent Virtual Reality (VR) is an emerging technology that enables users to partake in 360-degree virtual experiences using VR head-mounted displays \cite{kolesnichenko2019understanding, mcveigh2019shaping, mcveigh2018s}. VR offers full-body tracking and synchronous voice chat and has controllers that provide haptic feedback \cite{garrido2023sok}, allowing people to interact in newer, more immersive ways compared to traditional social media \cite{blackwell2019harassment}. While VR presents these novel affordances, it also lowers the bar for unwanted behavior by malicious social actors. The anonymity it provides to users~\cite{kim2023anonymity, duggan2017broader}, as well as the lack of their physical presence, not only increases the likelihood of harassment but also makes identification of harassers challenging~\cite{burke2015exploring}. While online harassment is not a new issue, the unique sense of embodiment and presence that VR enables \cite{blackwell2019harassment,adams2018ethics}, even without haptic technology \cite{qingxiao2022facing}, may amplify certain forms of harassment (e.g., sexual harassment) in virtual spaces.

Harassment in VR is becoming prominent today \cite{will2021grope,belamire2016,frenkel2021,unknown2018,unknown2022,duranske2007}, with an abusive incident estimated to occur every seven minutes \cite{metaverse7}. VR-based harassment may include virtual violence, virtual groping \cite{zheng2023safetyrisks}, and haptic sex crimes \cite{lemley2017law}. To enable users to deal with harassment, VR applications have introduced safety\footnote{We consider ``safety" in the context of interactions (with users, content, etc.) within the VR environment that threaten a user. We don’t consider physical safety implications arising intrinsically from VR hardware (e.g., headset, display), such as flashing lights causing epileptic seizures~\cite{lemley2017law, tychsen2020concern}, or injuries arising from careless use of VR (e.g., hitting a wall).} controls such as the personal bubble, power gesture \cite{Stanton2016}, safe zone \cite{meta2022}, etc. Targets of VR-based harassment have reported challenges in escaping and reporting problematic users~\cite{blackwell2019harassment}, highlighting limitations of existing safety controls. The set of safety controls is not standardized across VR apps, with high variance in functionalities they provide~\cite{zheng2023safetyrisks}, hinting at different usability challenges that may exist for even the same control across VR apps.

To identify gaps in our understanding of harassment controls and better address harassment in VR, it is essential to understand VR safety controls more deeply through the lens of targets subjected to VR-based harassment. First-hand accounts of these targets would elucidate the types of harassing activities that have a lasting impact and how the availability of safety controls, or lack thereof, contributed to these experiences. Further, it would enable understanding the usability and effectiveness of current safety controls and how they may be improved. When users' perspectives are contrasted with those of VR developers, i.e., those involved in the implementation of these features, the combined knowledge will highlight the gaps in practically addressing safety in VR.

Although prior works have identified types of VR-based harassment~\cite{freeman2022disturbing, blackwell2019harassment, zheng2023safetyrisks} and availability of safety controls~\cite{zheng2023safetyrisks} in VR, they have not studied the \emph{effectiveness} of safety controls and reporting mechanisms, nor have they investigated users' perceptions of their \emph{usability} at the time of harassment. We address this research gap by conducting semi-structured interviews with targets (n = $\NumberOfUserInterviews$) of VR-based harassment. We also augment our findings by interviewing (n = $\NumberOfDevInterviews$) VR developers to understand their perceptions on designing and deploying proposed safety features for VR. Specifically, we seek to answer the following research questions:

\noindent \researchquestion{1}{\xspace \textit{How do targets of VR-based harassment perceive the usability and effectiveness of existing safety controls and reporting mechanisms?}} We delve into the participants' thought processes behind (not) using safety controls and how such actions contributed to their experience of VR-based harassment.\\
\researchquestion{2}{\xspace \textit{What are the expectations and recommendations by targets of VR-based harassment for making VR safer?}} We understand what new safety measures our participants desire based on their experiences of VR-based harassment. \\
\researchquestion{3}{\xspace \textit{What are VR developers' perceptions of the design and deployment of safety controls?}} We understand the challenges VR developers perceive in implementing VR safety controls. Further, we understand developers' views on the feasibility of the safety features desired by targets of VR-based harassment. 

We performed a \emph{qualitative analysis} \cite{boyatzis1998transforming} of our participants' responses using open coding \cite{strauss1990basics}. In this paper, we make the following contributions:
\begin{itemize} 
    \item To the best of our knowledge, we are the first to conduct a multi-perspective study on VR safety through the lens of \textit{targets of VR-based harassment} and \textit{VR developers}. 
    
    \item We identify contexts where existing VR safety controls and moderation practices are non-usable and ineffective. For instance, VR users face usability challenges in finding users in crowded virtual spaces to block them. Safety controls are also ineffective in providing feedback to the harassers.
    
    \item We highlight VR users' expectations for making VR safer and contrast them with technical, legal, and financial challenges that VR developers perceive in implementing them. Users desire live moderation in social spaces and want users' behavior to be tracked across VR apps; however, VR developers highlight difficulties in deploying live moderation at scale and the privacy risks in tracking users. 
    
    \item We use our findings from this multi-perspective study to make recommendations to VR platform owners, app developers, and policymakers for improving safety in VR.
\end{itemize}

%% file: sections/02-relwork.tex
\section{Background \& Related Work}
\label{sec:relwork}

\paragraphb{Online Abuse.} Online harassment refers to behaviors that threaten or offend individuals through emails, instant messages, social media, etc.~\cite{hazelwood2013cyber,southworth2007intimate}. The ability of users to retain their anonymity~\cite{lapidot2012effects} exacerbates harassment, and the lack of their physical presence on these platforms makes their identification challenging, causing emotional distress to victims~\cite{burke2015exploring}.  

Prior works have extensively studied online abuse in non-VR contexts. Thomas et al.~\cite{thomas2021sok} created a taxonomy of types of online hate and harassment, identifying seven classes of attacks, including toxic content, content leakage, and impersonation, based on attackers' intents and capabilities. Of these, toxic content (e.g., bullying, trolling, hate speech) is viewed as the highest priority threat among experts as it incurs significant emotional harm~\cite{wei2023there}. Additionally, what constitutes toxic content differs across demographics, beliefs, and personal experiences~\cite{kumar2021designing}. Researchers have studied technology-based abuse specific to at-risk populations such as youth~\cite{freed2023understanding} and sex workers~\cite{bhalerao2022ethics, mcdonald2021s, strohmayer2019technologies}, as well as how technology facilitates intimate partner abuse and surveillance via mobile~\cite{almansoori2022global} and IoT devices~\cite{stephenson2023s, stephenson2023abuse}. Women, racial/cultural minorities, LGBTQ individuals, and persons with disabilities face more significant risks of online harassment \cite{fox2017women,mclean2019female}.

\input{tables/safety_controls}

\paragraphb{Safety in VR.} 
VR can act as a medium for harassing activities such as virtual violence, virtual groping \cite{zheng2023safetyrisks}, and haptic sex crimes \cite{lemley2017law}. The sense of embodiment and presence facilitated by 3D avatars in VR environments \cite{blackwell2019harassment,adams2018ethics}, and the ability of some users to feel as though they are their avatars even without haptic technology (termed as \emph{phantom sense}) \cite{qingxiao2022facing}, makes harassment in VR realistic and thus traumatizing. While more female avatars report VR harassment~\cite{shriram2017all}, vulnerable populations such as minors face risks of harm through virtual grooming and erotic role-play abuse~\cite{289530}.

\paragraphb{VR Apps \& Safety Controls.}
VR apps can be downloaded into standalone or tethered VR headsets \cite{angelov2020modern} through app stores such as Oculus Store \cite{oculus_store}, Steam VR \cite{steam_vr}, and Sidequest \cite{sidequest}. While every app has a primary purpose, such as watching movies  (e.g., `BigScreen'~\cite{bigscreen}) or competitive gaming (e.g., `Pavlov VR'~\cite{pavlovvr}), many apps have capabilities for multi-user interaction through social spaces and lobbies. Additionally, certain VR apps are primarily meant for social interaction, called Social VR apps (e.g., `VRChat' \cite{vrchat}).

VR apps have introduced safety controls such as muting, blocking, personal bubble and power gesture~\cite{Stanton2016}, personal boundary~\cite{oculus2022}, and safe zone~\cite{meta2022} to enable users to deal with harassment. Table \ref{table:safety_controls} describes the most prominent safety controls. Muting may be used either to disable one's own voice such that it cannot be heard by other users or disable other users' voices in a VR environment. "Blocking" --- also known as "ghosting" --- may make another user's avatar disappear or change its appearance, depending on the implementation. Proximity settings, also referred to as the "space bubble," "personal bubble," or the "personal boundary," enables a user to control the distance at which other users in a VR space can interact with them. "Safe zone" is a user's private space that can be accessed only by the user and may be invoked to "move away" from other users to a safe space. While the "safe zone" teleports the user to a pre-determined location/VR space, "quick travel" can be used to move to other VR spaces within the same app (not just the safe zone). Some apps have custom implementations, like in the power gesture case, which involves "putting your hands together, pulling both triggers and pulling them apart as if you are creating a force field"~\cite{Stanton2016}. Some VR apps have additional safety features such as `Safety and Trust System' (`VRChat') \cite{vrchat_safety}, `Comfort and Safety' (`RecRoom') \cite{recroom_safety}, etc. 

Safety controls may be used proactively (in anticipation of harassment) or reactively (after the harassment incident). Some are inherently used reactively (e.g., blocking, muting) while others (e.g., space bubble) are used proactively~\cite{freeman2022disturbing}. Additionally, VR apps offer ways for reporting harassment, which can be done inside the app, through the headset, or via email to the app developers. While controls like muting and blocking involve only users, reporting involves multiple stakeholders such as users, moderators and automated systems for toxicity detection~\cite{toxmod}.

Prior work has explored how users engage in social VR \cite{maloney2020falling} and uncovered tensions in specific interactions, such as between children and adults~\cite{maloney2020complicated,maloney2020virtual}. Moderating sensitive content~\cite{khlif2020virtual} and reporting users~\cite{deng2009users} are essential in dealing with inappropriate interactions in VR. Researchers have explored the ethics of acceptable behavior~\cite{sparrow2020silly,adams2018ethics}, and the influence of body-gender transfer in VR~\cite{doi:10.1177/1071181321651094}. 

Freeman et al. \cite{freeman2022disturbing} interviewed social VR users to understand the new characteristics of harassment emerging in social VR and also investigated users' strategies and recommendations to mitigate harassment. Blackwell et al. \cite{blackwell2019harassment} investigated users' expectations of social norms and moderation practices in VR communities. Zheng et al. \cite{zheng2023safetyrisks}  analyzed videos of VR activities posted by social VR users to identify types of safety risks in social VR, including virtual violence, abuse, and sexual harassment. Schulenberg et al.~\cite{schulenberg2023we} explored the (re)purposing of existing social VR features (e.g., boundary settings) for preventing interpersonal harm in VR. 

\paragraphb{Distinction from Prior Work.} 
While prior works have characterized harassment in VR and touched upon safety controls, they have not studied VR users' thought processes in (not) using specific safety controls or how the (non) usage of controls influenced their experience. To the best of our knowledge, we are the first to investigate the usability and effectiveness of safety controls through the lens of \textit{targets of VR-based harassment}. We further augment our findings with \textit{VR developers'} perspectives on designing and deploying VR safety measures.

%% file: tables/safety_controls.tex
\begin{table}[!t]
	\centering
 \caption{Prominent safety controls available in VR apps.}
 \label{table:safety_controls}
 \resizebox{0.95\linewidth}{!}{
		\begin{tabular}{ll} 
			\toprule 
			 \multirow{1}{*}{\textbf{Safety}} & \multicolumn{1}{c}{\textbf{Function}} \\
            \textbf{Control} & 
			\\ \midrule
			 Mute & Disable voice chat of self, or other users in a VR space \\
             \midrule
			Block & Hide or change the appearance of user(s) in a VR space\\
             \midrule
			Proximity & Control the distance at which other users can interact\\
              setting & with a user in a VR space 
            \\ \midrule
			\multirow{1}{*}{Quick travel} & Travel to a different location within a VR app \\
            \midrule
			\multirow{1}{*}{Safe zone} & A user's private space accessible only to that user\\
            \midrule
			\multirow{1}{*}{Vote kick} & Kick a user out of a VR space based on majority vote\\
            \midrule
			\multirow{1}{*}{Trust rank} & Levels of trust assigned to a user
			\\ \bottomrule
		\end{tabular}}
  \vspace{8pt}
\end{table}

%% file: sections/03-method.tex
\section{Methods}
\label{sec:method}

\noindent We conducted a phased multi-perspective study with \NumberOfUserInterviews targets of VR-based harassment (\textbf{Study-I}) and \NumberOfDevInterviews VR developers (\textbf{Study-II}). The findings from \textbf{Study-I} were used to design parts of \textbf{Study-II}. Table~\ref{tab:methods_table} details our recruitment and interview procedure for both studies. We further highlight our study design, data analysis methods, and limitations.

\input{tables/methods_table}

\subsection{Study Design \& Ethical Considerations}

\noindent For \textbf{Study-I}, we advertised the study in VR-specific online forums. Apart from the recruitment platforms specified in Table~\ref{tab:methods_table}, we circulated our study in forums for women in VR (e.g., Ladies of Population One) to recruit participants representing marginalized groups in tech spaces~\cite{blackwell2017classification, collins2017tech, us2018employed}. A few participants we interviewed shared the study with other groups, facilitating snowball sampling~\cite{parker2019snowball} (four participants were from groups where we did not directly advertise). We selected a stratified sample of participants for interviews based on their responses to the screening survey (looking at VR usage and harassment types experienced), irrespective of where they obtained the study information.

We consulted a psychology expert during study design\footnote{We provide all study materials at: \href{https://doi.org/10.17605/osf.io/c7fks}{\textbf{https://doi.org/10.17605/osf.io/c7fks}}}~\cite{bellini2023sok} and determined to exclude those diagnosed with post-traumatic stress disorder (PTSD)~\cite{bisson2007psychological} or seeking help for emotional distress. The expert reviewed our interview script and advised us on phrasing sensitive questions without triggering individuals. In our screening survey, we stated that selected participants would be asked about their harassment experience in VR, but also specified that talking about traumatic events can be cathartic~\cite{cathartic}. We curated a list of mental health resources in case of emotional distress during the interview, which included links to websites containing details about mental health hotline numbers to institutions such as the Center for Mental Health Services.

During the interview, in order to study usability challenges in VR safety controls, we first collected contextual information: we asked participants to specify the VR apps in which they experienced harassment (we focused on four apps utmost), then probed about their use of safety controls and reporting mechanisms (explaining the terms when necessary) during/after the incident, and finally asked about what they desired for improving VR safety. After the interview, we inquired about the participants’ emotional state, and none reported adverse effects. Regardless, we shared mental health resources with all the participants. Participants were given the opportunity to review the interview transcripts; 11 reviewed them, reporting accurate captioning.

We developed \textbf{Study-II} to contrast the findings of \textbf{Study-I} with developers’ perspectives by identifying developmental challenges and practical solutions. Our screening survey for \textbf{Study-II} asked about participants' roles as VR developers, the apps they developed, demographic information, and their LinkedIn profiles to ascertain their fit for the study. Further, the survey had questions about whether participants had developed apps containing features like socializing, multiplayer gaming, learning, or streaming. We tried to prioritize those with experience in social and multiplayer categories (since those categories of apps have more opportunities for harassment) but eventually included others, too.

We started the interviews by asking participants about their VR development experience and their perspective on handling harassment in VR. Then, we asked about the app development pipeline they were involved in and the challenges they perceived in implementing safety features. We also gathered their perspectives on the feasibility of users’ desired features. Our protocol also included questions about reporting and moderation. However, none of our participants were content/user moderators thus we could not report related findings. Questions about the developers' experiences of VR-based harassment were not part of the interview protocol, but some participants shared their experiences.

Both \textbf{Study-I} and \textbf{Study-II} were approved by our Institutional Review Board (IRB). All audio recordings were transcribed and de-identified immediately after the interviews. Participants also had the option to withdraw from the interview at any time. 

\subsection{Data Collection and Analysis}

\noindent We hosted the surveys of our study (screening, demographics, compensation) using Qualtrics \cite{qualtrics2020}. We refined our interview protocol by piloting with \NumberOfPilotinterviews participants for \textbf{Study-I} and \NumberOfDevPilotinterviews for \textbf{Study-II}. Participants of both studies were given a \InterviewCompensation Amazon gift card after they completed the interviews. All the interviews were audio-recorded and transcribed using Whisper~\cite{radford2022robust}. One researcher checked each transcript for accuracy. Two researchers conducted thematic analysis~\cite{boyatzis1998transforming, strauss1990basics} on the interview transcripts by independently coding the transcripts for both studies. For \textbf{Study-I}, the transcripts were divided into four sections (harassment incident, use of safety controls, use of reporting, expectations for safer VR). Each section was coded for all transcripts (in batches of five), followed by discussions to generate/update the codebook before moving on to the next. In the case of \textbf{Study-II}, the researchers coded two transcripts together to create the initial codebook, and the rest of the transcripts were coded independently (in batches of two) and discussed for updating the codebook. For both studies, multiple discussions were conducted to reach an agreement to generate a codebook. Since our coding process involved multiple iterations and discussions, intercoder reliability was not necessary to be checked~\cite{mcdonald2019reliability}.

\input{tables/taxonomy_of_harassment}

\subsection{Limitations}

\noindent As is typical with interview studies, our recruited sample size was relatively small due to the sensitive topic of study (\textbf{Study-I}) and challenges in recruiting VR developers (\textbf{Study- II}). While the exclusion criteria for \textbf{Study-I} prevented us from interviewing those impacted deeply by VR-based harassment, we minimized risks to participants. To preserve participants' anonymity while sharing sensitive experiences of harassment, we refrained from having identity verification. However, after the interviews, we identified one imposter participant~\cite{roehl2022imposter} who participated twice, based on the high similarity of responses\footnote{We initially conducted 20 interviews and later excluded the imposter participant's data, totaling the number of valid participants to \NumberOfUserInterviews.}. For \textbf{Study-II}, we required participants to provide their LinkedIn profile during screening and enable their webcams prior to the interview\footnote{When one of the participants failed to enable their camera at the start of the interview in Study-II, the researcher did not proceed with the interview.}. In \textbf{Study-II}, most developers we interviewed were from small teams, which may not exactly represent development experiences for platforms our participants interacted with. However, safety is relevant in all apps, even those developed by smaller teams. Both studies relied on self-reported information from participants, which may be subject to social desirability bias. To address these limitations in future research, participants may be recruited from a broader range of forums, languages, and cultural backgrounds.

%% file: tables/methods_table.tex
\begin{table*}[!t]
	\centering
 \caption{An overview of methods for \textbf{Study-I} (with targets of VR-based harassment) and \textbf{Study-II} (with VR developers).}
 \label{tab:methods_table}
 \resizebox{1.0\textwidth}{!}{
		\begin{tabular}{lll} 
			\toprule 
			  & \multicolumn{1}{c}{\textbf{Study-I}} & \multicolumn{1}{c}{\textbf{Study-II}}\\
        \midrule
        {\footnotesize Recruitment} & {\footnotesize VR-specific Discord servers, subreddits, Facebook groups} & {\footnotesize VR-specific LinkedIn groups, subreddits}\\
        
        \midrule

        {\footnotesize Inclusion} & {\footnotesize Ages 18-64, resident of the US, active user of VR, experienced} & {\footnotesize Ages 18 and above, resident of the US, VR developer}\\

        {\footnotesize Criteria} & {\footnotesize  harassment in VR, not diagnosed with PTSD, not seeking} & {\footnotesize }\\

        {\footnotesize } & {\footnotesize help for emotional distress at the time of the study} & {\footnotesize }\\
        
        \midrule

        {\footnotesize Screening} & {\footnotesize Type of harassment experienced in VR (trolling, bullying, etc.} & {\footnotesize Roles held as VR developer (UI/UX designer, XR gameplay and tools }\\

        {\footnotesize Questions} & {\footnotesize  presented as a list of categories), VR apps in which harassment } & {\footnotesize  engineer, AR/VR maintenance and support, etc.), details of VR apps}\\

        {\footnotesize } & {\footnotesize  was experienced, VR usage} & {\footnotesize currently developing}\\        
       
        \midrule

        {\footnotesize Demographics} & {\footnotesize Just before the interview, to minimize sensitive data collection} & {\footnotesize Part of the screening questionnaire}\\

        {\footnotesize Collection} & {\footnotesize at screening phase (as participants were targets of harassment)} & {\footnotesize }\\

        \midrule

        {\footnotesize Interviews } & {\footnotesize 18 targets of VR-based harassment, conducted via Zoom from } & {\footnotesize 9 VR developers, conducted via Zoom from August 2023 to  September}\\

        & {\footnotesize November 2022 to February 2023, lasting 59 min on average} & {\footnotesize  2023, lasting 49 min on average} \\
        
        \midrule

        {\footnotesize Participant } & {\footnotesize  \NumberOfFemale female, \NumberOfNonbinary non-binary, \NumberOfNoncis non-cis, and \NumberOfBlack black persons;  3} & {\footnotesize 6 professional developers, 2 VR-based researchers, 1 hobby developer,  }\\

        {\footnotesize Distribution} & {\footnotesize participants were prominent in certain VR communities (T12, } & {\footnotesize including the founder of a VR game and a doctorate in XR. }\\
        
        {\footnotesize } & {\footnotesize   T13, T15). Refer Table~\ref{table:demographics} Appendix~\ref{sec:user_demographics} for more details.} & {\footnotesize Refer Table~\ref{table:dev_demographics} in Appendix~\ref{sec:dev_demographics} for more details.}\\
        
        \midrule

        {\footnotesize Interview } & {\footnotesize  (1) Describe harassment incident experienced in VR,  } & {\footnotesize (1) VR development experience, (2) Solving usability challenges }\\

        {\footnotesize Protocol} & {\footnotesize (2) Use of safety controls, (3) Use of reporting mechanisms,  } & {\footnotesize  in VR safety controls, (3) Challenges in implementing VR safety features,}\\
        
        {\footnotesize } & {\footnotesize (4) Expectations for enhancing safety in VR} & {\footnotesize (4) Feasibility of user recommended features}\\

    \bottomrule	
\end{tabular}}
\vspace{-8pt}
\end{table*}

%% file: tables/taxonomy_of_harassment.tex
\begin{table*}[!t]
	\centering
 \caption{Harassment incidents reported by the participants of \textbf{Study-I}. Each harassment incident is categorized according to taxonomies from prior work on online abuse attacks by Thomas et al.~\cite{thomas2021sok} and VR safety risks by Zheng et al.~\cite{zheng2023safetyrisks}.}
 \label{tab:taxonomy_of_harassment}
 \resizebox{1.0\textwidth}{!}{
		\begin{tabular}{ccl} 
			\toprule 
			 \multirow{1}{*}{\textbf{\footnotesize Harassment}} & \multicolumn{1}{c}{\multirow{2}{*}{\textbf{\footnotesize Participants}}} & \multicolumn{1}{c}{\multirow{2}{*}{\textbf{\footnotesize Excerpts from participants}}}\\
         \multirow{1}{*}{\textbf{\footnotesize Type}} & & \\
        \midrule

	    \multirow{2}{*}{\footnotesize Trolling /}  & \multirow{1}{*}{\footnotesize T3, T4, T7, T8, T10, T13,} & 
        \textit{\footnotesize "My accent is different than the average American, and most of the players in this game  }\\
        \multirow{2}{*}{\footnotesize Virtual abuse} & \multirow{1}{*}{\footnotesize  T14, T15, T16, T17, T18} & \textit{\footnotesize are either American or British, I’ve had players chase me around calling me f*gg*t or }\\ 
        & \multirow{1}{*}{\footnotesize } & \textit{\footnotesize gay or little b*tch the entire match." (T13, Echo VR)}\\ 

        \midrule
   
        \multirow{1}{*}{\footnotesize Profanity /} & \multirow{1}{*}{\footnotesize T1, T2, T3, T4, T5, T9,} 
        & \textit{\footnotesize "We're doing a team task or something like that. They just, started saying some  }\\
        \multirow{1}{*}{\footnotesize Virtual abuse} &  \multirow{1}{*}{\footnotesize T10, T14, T15 T16}
        & \textit{\footnotesize inappropriate vulgar things and I didn't really feel comfortable. (T1, Asgard's Wrath)"} \\
        
        \midrule
   
        \multirow{1}{*}{\footnotesize Hate speech /} &  \multirow{2}{*}{\footnotesize T5, T8} 
        & \textit{\footnotesize  "He would just yell at me. I think he must have had anger management issues or something; } \\
        \multirow{1}{*}{\footnotesize Virtual abuse} & \multirow{3}{*}{\small } 
        & \textit{\footnotesize he would say random over-the-top vitriolic things. He sounded furious." (T8, Pavlov VR)} \\

        \midrule
   
        \multirow{1}{*}{\footnotesize Threats} & \multirow{2}{*}{\footnotesize T16} 
        & \textit{\footnotesize "He started saying, 'I am going to gut your mother and skin your family,' kept repeating it  }\\
        \multirow{1}{*}{\footnotesize of violence}  & & \textit{\footnotesize over and over. That was a bit scary, like, really unsettling." (T16, Echo VR)} \\

        \midrule        
        
        \multirow{1}{*}{\footnotesize Bullying /} & \multirow{1}{*}{\footnotesize T3, T7, T8, T9, T11, T13, } 
        & \textit{\footnotesize "In the game, you get punched in the head, and it stuns you. I've had players chase me  } \\
        \multirow{1}{*}{\footnotesize Virtual violence} & \multirow{1}{*}{\footnotesize  T14, T15, T17, T18} 
        & \textit{\footnotesize  around, trying to punch me in the head the entire match." (T13, Echo Arena)} \\

        \midrule
   
        \multirow{2}{*}{\footnotesize (Virtual) Sexual} & \multirow{3}{*}{\footnotesize T4, T7, T12, T13} 
        & \textit{\footnotesize "There would be a guy that starts to pretend to r*pe me and encourages others. Because }\\
        \multirow{2}{*}{\footnotesize harassment} & \multirow{3}{*}{\footnotesize }  &
        \textit{\footnotesize there's presence in VR, they're doing this physically to my avatar, putting their avatar's }\\
        \multirow{3}{*}{\footnotesize } & \multirow{3}{*}{\footnotesize }  &
        \textit{\footnotesize     cr*tch in my face, making slurping noises, sucking [my] b**bs." (T12, Echo VR)}\\
        
        \midrule
   
        \multirow{1}{*}{\footnotesize Explicit content} & \multirow{1}{*}{\footnotesize T2, T6, T11} 
        & \textit{\footnotesize "Seeing content, I'm not okay with, people not fully dressed." (T6, YouTube VR)}\\

        \midrule
   
        \multirow{1}{*}{\footnotesize Virtual crashing} & \multirow{1}{*}{\footnotesize T11} 
        & \textit{\footnotesize "I’ve played it before. Someone had toyed with that game." (T11, House of Terror)}\\
        
        \midrule
   
        \multirow{1}{*}{\footnotesize Virtual trash} & \multirow{2}{*}{\footnotesize T14} 
        & \textit{\footnotesize "You can high-five each other. You’ll see people slap your hand a bunch of times to get }\\
        \multirow{1}{*}{\footnotesize actions } &  {}
        & \textit{\footnotesize  the high five. When you don't, they’ll just wave in front of your hand." (T14, BigScreen)} \\

          \midrule
   
        \multirow{1}{*}{\footnotesize Misuse of} & \multirow{3}{*}{\footnotesize T8} 
        & \textit{\footnotesize "Because everybody looks the same, if he found out a specific person was me, he would }\\
        \multirow{1}{*}{\footnotesize safety features} &  {}
        & \textit{\footnotesize team kill." (T8, Pavlov VR)} \\

    \bottomrule	
\end{tabular}}
\vspace{-8pt}
\end{table*}

%% file: sections/04-results-01.tex
\section{Users' Perceptions on VR Safety}
\label{sec:effectiveness}

\noindent 
In this section, we describe our findings from \textbf{Study-I} (\S~\ref{sec:method}), outlining results from every part of our interview protocol. In the first section of the interview, we asked participants about the harassment experiences they had in VR apps to contextualize their use of safety controls (\S~\ref{sec:harassment}). In the second part of the interview, we focused on participants' awareness of safety controls available in VR apps, as well as their perceptions of the usability and effectiveness of safety controls (\S~\ref{sec:perceptions_safety_controls}). In the third part, we studied the effectiveness of reporting mechanisms (\S~\ref{sec:perceptions_reporting}). Finally, we asked about their expectations for enhancing safety in VR(\S~\ref{sec:user_expectations}). Our participants' diversity (Table~\ref{table:demographics} in Appendix~\ref{sec:user_demographics}) enabled us to capture varying perspectives stemming from a wide range of VR experiences. 

\subsection{Harassment Experienced by Participants}
\label{sec:harassment}

\input{tables/source_of_harassment}

\noindent  We asked participants to specify the VR apps in which they experienced harassment (Table~\ref{table:demographics} in Appendix~\ref{sec:user_demographics} lists the apps). Overall, we consider \ParticipantVRAppPairs participant-app pairs from \NumberOfUserInterviews participants across \SocialVRApps social VR, \GamingVRApps gaming VR, and \StreamingVRApps streaming VR apps. We categorized the harassment incidents reported by participants according to existing taxonomies on online abuse~\cite{thomas2021sok} (trolling, profanity, hate speech, etc.) and VR safety~\cite{zheng2023safetyrisks} (virtual violence, virtual crashing, virtual sexual harassment, etc.). Table~\ref{tab:taxonomy_of_harassment} lists example quotes for each type of harassment recounted by our participants. Here, we do not introduce new findings about harassment in VR; rather, we illustrate the different harassment types to set the context for discussing safety features in the following sections.

We identified that most of our participants experienced different forms of toxic content, apart from VR-specific scenarios such as virtual crashing (using tactics or bugs to ruin others’ experience) and trash actions (activities typically intended to spoil the experiences of others)~\cite{zheng2023safetyrisks}. Participants described their experiences as \textit{`annoying'}, \textit{`uncomfortable'}, \textit{`violating'}, and \textit{`jarring'}. Participants reported experiencing harassment based on their \emph{gender} (as in \cite{shriram2017all, freeman2022disturbing}), with five out of six of our female participants reporting gender-based harassment by a male-like avatar. They also reported that \emph{race}, unique \emph{avatar attributes}, and \emph{physical attributes} that helped identify a user's gender \cite{freeman2022disturbing, zheng2023safetyrisks}, contributed to harassment, as illustrated in Table~\ref{tab:source_of_harassment}.

In summary, although many accounts of VR-based harassment can be categorized as "toxic content"~\cite{thomas2021sok}, VR amplifies many forms of abuse (e.g., virtual sexual harassment, bullying) due to its immersion and presence~\cite{zheng2023safetyrisks} when compared to traditional social spaces online. These incidents leave a significant psychological impact, as reported by our participants. 

\subsection{Perceptions on Safety Controls}
\label{sec:perceptions_safety_controls}

\noindent After asking about the harassment incident experienced by the participant, we asked them if they were aware of safety controls at the time of harassment and if they used them. Based on their usage, we asked them about the usability and effectiveness of the safety controls they used. The list of safety controls was not picked beforehand; participants came up with controls themselves during the interview.

\paragraphb{Awareness of Safety Controls.}
Of the \ParticipantVRAppPairs participant-app pairs considered, in \NumberAwareOfControls of the cases, participants were aware of at least one safety control at the time of harassment, which included muting (\NumberOfMute), blocking (\NumberOfBlock) and ghosting (\NumberOfGhost), using safety bubble (\NumberOfSafetyBubble), vote kicking (\NumberOfVoteKick), a form of teleportation such as quick travel (\NumberOfQuickTravel) or changing lobbies (\NumberOfChangingLobbies). In addition, one participant reported using parental guidance in `YouTube VR' as a safety control. Participants learned about safety controls in a variety of ways. In \NumberOfTutorial of the cases, they were informed because of a tutorial available on the VR app. In \NumberOfNudges cases, the nudges in the app (`Echo VR') prompted them to learn about the controls. In \NumberOfOtherGames cases, their knowledge was due to contextual information from playing other video games. Participants also discovered the controls by chance, through exploring the app or searching the Internet.

However, the participants who were unaware of safety controls before encountering harassment did not necessarily attempt to learn about them after the incident:
\begin{myquote}
"No, absolutely not. I just left it. [Using the app] is like a fun thing for me to do. I can just find another platform to just move on to." (T1, Asgard's Wrath)
\end{myquote}
T14 added that learning the safety controls was not worth his time and hinted at an unintuitive design:\textit{"VRChat is a complicated app. I don't know how it works. It's not, to me, user-intuitive. The offenses are so prevalent that it is not worth my time putting in the time to learn"}.

Some participants (T4, T11) reported having checked for controls yet did not make changes to their settings. While T4 did not understand the controls, T11 felt that the trauma of his experience was too much for him to continue using VR.

\paragraphb{Activation of Safety Controls.}
Participants used safety controls both proactively and reactively~\cite{freeman2022disturbing}, depending on previous experience of harassment and when they learned about them. Muting was the most predominantly used control (\NumberOfEnabledMute), followed by blocking (\NumberOfEnabledBlock) and ghosting (\NumberOfEnabledGhost). The use of quick travel, vote kicking, and parental guidance was mentioned by one participant each. Some controls had several flavors based on the context; for instance, "mute" could be "mute self," "mute others," or "mute all." 

T5 talked about the dependence of certain VR games (e.g., `Beat Saber') on external platforms (Discord) and how controls from those platforms need to be used to mute people while playing the VR game:\textit{
"Beat Saber is multiplayer, but there's no voice chat in the game. You have to rely on Discord servers to get into a multiplayer lobby with players, [but there] it's mostly 10-year-olds saying [the] N-word."}

Some participants had clear preferences for using one safety control over another. T5 preferred blocking to muting: \textit{"I'm not going to disable the voice chat in-game [Beat Saber] because it's not everyone [who harasses]. It's only a few people, and blocking them usually helps"}. T8 described his preference for vote kicking in `Pavlov VR':\textit{"You can mute and also vote kick people. You can call a vote, and if it gets enough votes, the person is kicked from the lobby. It is a necessity to hear everything people are saying, so I don't mute."}

Of the \NumberAwareOfControls cases where participants were aware of the safety controls in the apps, in \NumberOfAwareAndEnabled of them, they were enabled, while in \NumberOfAwareYetNotEnabled, they were not. T12 did not use safety controls as she did not think they stopped the harassment:
\begin{myquote}
    "If you mute or block them, it's not going to stop the harassment. It's just going to stop me from being aware of it. If they stick their cr*tch in my face, others can see that even if I can't." (T12, Echo VR)
\end{myquote}
T17 added that safety controls were implemented only in the lobbies and did not stop harassment while playing `Echo VR'. T8 found that it was faster to leave `VRChat' than to individually block every offender.

\paragraphb{Usability of Safety Controls.}
Among \NumberAwareOfControls instances where participants were aware of safety controls and \NumberAwareOfControlsLater instance where they learned about them after a harassment incident, in \NumberOfEasy cases, they found the process of enabling the controls to be easy. In \NumberOfNotEasy cases, they found it to be cumbersome and challenging. Moreover, individuals' perceptions of ease of use varied considerably for the same set of features in a VR app. For example, T15 felt: \textit{"The in-game [Echo VR] features to mute or ghost are very simple. You bring up your menu and select the individual or choose the easier option to mute or ghost all"}.
In contrast, T12, also a long-term user of `Echo VR,' said:\textit{"There is a block, but it's hard to use because you have to be able to point at the avatar, which is difficult sometimes"}. T16 recalled how it was hard to find the mute button on `RecRoom' as a new user:
 \begin{myquote}
      "You had to go into some sub-menu, [with] all the people in the lobby listed, and find their name and click on it. The names are not always super easy to see if they're moving around." (T16, RecRoom)
 \end{myquote}

Participants also noted that specific controls were more usable than others. T8 found it easier to mute himself than to mute others on `VRChat':\textit{"It's easy to mute yourself. So that's what I ended up doing most of the time. Blocking people isn't necessarily difficult, but not as fast"}. 

\paragraphb{Effectiveness of Safety Controls.}
Participants found safety controls to be useful in certain situations, and we find that the effectiveness of safety controls is highly context-dependent. In social VR, users perceive muting to effectively filter out inappropriate comments and overcome verbal disruption. In gaming VR, which shares the culture of trash-talking with other forms of online gaming \cite{cote2017can}, users state that muting is effective. T16, who muted his offensive team player on `Echo VR', said:\textit{"Even if the person said horrible things, I assume they still want to win the game. So they're not going to come over and intentionally play badly"}. 

Participants used blocking to effectively limit further interaction with the harassers, especially when there were only a few of them to deal with. They also found quick travel to be effective in escaping from the harasser:
\begin{myquote}
  "They don't know where you're going, so they can't chase you around." (T7, Zenith) 
\end{myquote}

\paragraphb{Ineffectiveness of Safety Controls.}
Although participants found the use of safety controls effective in a few scenarios, in a vast majority of harassment incidents, they found the controls lacking in several ways. Social VR users felt that safety controls affected social interactions with non-harassers. T16 noted that enabling safety controls \textit{did not provide feedback to the harassers}, and thus did not stop their behavior:
\begin{myquote}
    "I don't think that the game tells them that I have muted them. So they would have no feedback [that] this person can't hear me." (T16, RecRoom)
\end{myquote}

T7 complained that blocking did not remove the harasser from the game and only changed their appearance on `Orbus'. She added that quick travel disrupted her experience on `Zenith':\textit{"You're doing something, and you have to stop what you're doing. It's like you're being punished because you're the one being harassed"}. T5, who used `Beat Saber' and connected to Discord for voice chat, explained how blocking users on the game did not mute them on Discord and had to use the safety controls on multiple platforms: \textit{
"Because blocking the person doesn't mute the person on Discord, I'll just mute the person"}. T13 expressed how certain safety controls would affect communication, which was essential in a strategic game like `Echo VR':\textit{"Muting players during the game makes it harder because it's a team game, and you can't communicate if you're muting people"}. T9 felt that none of the safety controls would be effective as the harasser could simply create a new account on the app: \textit{"If the guy creates another profile, the blocking and muting would be in vain. You'd have to block this new avatar now. Back to square one"}. T13 and T18 also felt that safety controls would not stop the culture of abuse on `Echo VR':
\begin{myquote}
    "They have zero effect on the larger culture of abuse. They're a testament to the failure of the larger structure to have any form of protection or any meaningful way to stop or intervene in whatever negative dynamics are going on." (T18, Echo VR)
\end{myquote}
T8 explained how safety controls could be misused, as described in Table~\ref{tab:taxonomy_of_harassment}. T14 felt that the use of safety controls could not prevent disruption when a new user joined the room on `BigScreen':\textit{"If someone new comes in and doesn't have that person blocked, then you'll have someone being like, oh, who is the screecher? And everyone has to be like, it's this person, block them"}. T14 also felt that the safety controls were ineffective in the unique context of `BigScreen':
\begin{myquote}
    "You can pull up the usernames, but it'll be every username sitting in a theater. If you see someone screeching, you can see a microphone going off. But they'll be quiet when they're doing hand slapping. So you can't see who's talking. Sometimes they will get underneath the seats and slap your hand where you can't see their username. So everyone at that point has to stop the movie and find the offensive person." (T14, BigScreen)
\end{myquote}

\paragraphb{\underline{Takeaways:}} Muting is the most used safety control, in line with trolling and profanity, which are reported to be the most dominant types of harassment. Blocking and proximity settings that help mitigate bullying and sexual harassment are the second most frequently used controls. The key usability challenges with safety controls arise when selecting them from a dense hierarchy of menus while identifying the offending user's name from a long list of lengthy usernames (often with many special characters) or pointing at an offending user's avatar in order to take action on them, while they are still moving. Although safety controls provide a temporary escape to the target, they affect communication with non-harassers, fail to provide feedback to harassers upon muting or blocking, do not remove the harasser from the game but merely change their appearance, and could be misused to cause further harassment. Additionally, they fail to prevent others from witnessing the incident and stop further instances of harassment. 

\subsection{Perceptions on Reporting}
\label{sec:perceptions_reporting}

\noindent After asking about participants' experiences with safety controls, we also asked them about their experiences with using the reporting mechanisms in VR. Reports can be made in-app, via the headset, or through the website, depending on the VR app and platform in question. We were particularly interested in the reporting process as it is a multi-stakeholder process involving not only the users (unlike the rest of the safety controls) but also developers/moderators and, in some cases, automated toxicity detection systems. Since reporting often involves feedback to the reporting/reported users, we investigated participants' satisfaction with the reporting systems in the VR apps they used. 

We find that participants only submitted reports in half of the cases. Their failure to report was either due to a lack of reporting mechanisms, knowledge of reporting, or evidence. Some users did not want to spend effort on reporting. In cases where users submitted reports, their ease of reporting was dependent on the type of evidence they needed to provide.

\paragraphb{Why Users Do Not Report.} Of the \ParticipantVRAppPairs cases of harassment, only \NumberOfReported cases were reported. Several participants did not know how to report. T16 did not know the controls to report and preferred to leave the app (`RecRoom'), while T7 perceived reporting to be difficult and did not want to "mess" with the process of reporting:
\begin{myquote}
   "I'm not sure how you report someone in Zenith and also I just didn't feel like messing with it." (T7, Zenith)
\end{myquote}

\noindent T8 and T16 both mentioned that `Pavlov VR' did not have the feature to report and expressed their lack of faith in the developers to take necessary action. 
Some users who were new to the app assumed that the process would be cumbersome based on their experience with other apps. For example, T7 recalled her experience using the in-app keyboard on `Orbus' and assumed that the process would be similar on `Zenith':
\begin{myquote}
    "I'm sure it would be like Orbus, where you have to fill something out in the game on a keyboard. It's really cumbersome to do that. Plus, they were being very annoying, and I didn't want to stay long enough to do it." (T7, Zenith)
\end{myquote}

\noindent In a few cases, participants reported that it did not occur to them to report. Some participants also failed to report as they believed that the existing policies would not consider a certain type of harassment as a violation. For instance, T5 (`Beat Saber') felt: \textit{"You can't do anything against him because he's technically not breaking any rules"}. Several other participants echoed the notion that reporting did not have any effect. T18, who was banned for 24 hours on `Echo VR' due to a retaliatory report by his harasser, expressed outrage:
\begin{myquote}
    "That incensed me even more. I got banned, and it's only for 24 hours. You expect me to experience the negativity, to videotape the negativity, to go out of my way and submit the report to you. And then you're going to kick them out for 24 hours or 48 hours? How much of a punishment is that? That's a total waste of my time." (T18, Echo VR)
\end{myquote}
In some cases, even when the user knew how to report, they could not because they did not have evidence. T17 said: \textit{"You have to provide footage of the incident, and I didn't"}.

\paragraphb{Ease of Reporting.}
Our participants submitted reports through the in-app button, headset, or an external website affiliated with the VR app. Six participants reported that they found the reporting process to be simple and easy, and we note that in all of these cases, the only information they needed to provide was a description of the incident or choose the type of incident from a drop-down menu. 

T7 found using the in-game keyboard on `Orbus' for reporting cumbersome and frustrating. T17 felt that reporting was complex as she could not capture everything her harassers said before she started recording. T13 noted that reporting could be difficult for beginners, but it might get easier with practice. T14 added: \textit{"The only reason I know it so well is because I do it so much. I put time into the community, and I'm on all the Facebook and Discord groups. If a new user were to do it, it's not so clear how to report somebody [on Echo VR]. It's not part of the tutorial at all."}

Of the five cases in which the user had to provide a video of the incident, four participants found the process difficult.
\begin{myquote}
    "Each report takes 15 minutes if you're doing it properly. You have to get off the game, go through your footage, do a small edit of it, and write out the email. Unfortunately that's why many people don't do it." (T15, Echo VR)
\end{myquote}
T12 said that reporting could take five minutes to a week and highlighted some challenges with headset-based reporting: 
\begin{myquote}
    "On Quest 2, you click a report button, and you can record a snippet or upload a video. It'll search for the person, and you just send in the report, which is relatively easy. But you can't shorten it. You can't do that in the headset, you need to go to your computer." (T12, Echo VR)
\end{myquote}

\paragraphb{How Reporting was Handled.}
For 11 out of 16 reports, participants believed a moderator had processed their reports. In five of these cases, they were automated responses acknowledging the report. In some cases, the participants assumed that an action had been taken when they did not see the harasser's account anymore (T10, `Second Life') or did not encounter the type of content they reported anymore (T6 and T11, `YouTube VR'). Four participants received a response specifying whether action was taken.

In nine cases, participants were satisfied with the responses, while in the other two, they were not. T13 felt that the systems were underdeveloped and human resources were lacking:
\begin{myquote}
    "I reported to Meta, and I got an email saying that it does not breach their terms of service before I got an email saying, `we will review your complaint'. So, it's not exactly encouraging." (T13, Echo VR)
\end{myquote}
T14 added that while he was satisfied with the platform's actions, he expected more from the app developers:\textit{"With Meta, [I'm satisfied]. I think that more things should get action taken than does. But I wish Echo VR, the company, was taking action and didn't just pass it off to Meta"}.

T16, who did not hear back from the moderators of `Echo VR,' found reporting to be disincentivizing as he had \textit{"no idea whether it's actually doing anything"}. He believed it was essential for the feedback to include what action was taken and what abilities had been restricted or revoked for the harasser. He also suggested not including the harasser's username in the feedback in order to protect their privacy. T2 added that reports should be taken seriously, considering the target's mental health, while T3 wanted moderators to enquire about the well-being of the user who reported. 

\paragraphb{\underline{Takeaways:}} If users do not report, it may be due to a lack of reporting mechanisms, knowledge of reporting, or evidence. Although users might find reporting easy once they get acquainted with the system, some aspects of existing reporting mechanisms, such as entering text through a keyboard while in VR or capturing video evidence, are cumbersome. Moreover, users expect timely responses to their reports, with feedback on the action taken. 

%% file: tables/source_of_harassment.tex
\begin{table*}[!ht]
	\centering
 \caption{Sources of harassment identified by participants.}
 \label{tab:source_of_harassment}
 \resizebox{1.0\textwidth}{!}{
		\begin{tabular}{cll} 
			\toprule 
			 \textbf{Source} & \textbf{Participants} & \textbf{Excerpts from participants}\\
        \midrule
        
        \multirow{3}{*}{\small Gender} & \multirow{2}{*}{\small T3, T4, T7} 
        & \textit{\footnotesize "Oh, come over here, you b*tch, let me do blah, blah, blah. Women shouldn't even be in this game anyway.}\\
        & \multirow{2}{*}{\small T12, T17} 
        & \textit{\footnotesize Why don't you get off and go make me a sandwich". (T12, Echo VR) } \\
        &  & \textit{\footnotesize "It was really frustrating, women are made to be some kind of joke in VR." (T17, Echo VR)}\\ 
    
        \midrule
    
	    \multirow{2}{*}{\small Race}  & {\small T4, T13,} & 
        \textit{\footnotesize "I wanted to play a game with another user and the person wasn’t interested in playing with me }\\
        & {\small T15, T16} & \textit{\footnotesize because he was white and I was black." (T4, Beat Saber)}\\ 
        
        \midrule
   
        \multirow{3}{*}{\small Avatar} &  \multirow{3}{*}{\small T15, T16,} 
        & \textit{\footnotesize "One time I was wearing a turban and jeez, a bunch of kids walked up and wouldn’t leave me alone. They} \\
        \multirow{3}{*}{\small attributes} & \multirow{3}{*}{\small T9, T11} 
        & \textit{\footnotesize kept pestering me and asking if I was Arabic, or Muslim. I didn’t really know what to do." (T16, RecRoom)} \\

        & & \textit{\footnotesize "The special season avatar, it’s like a dog head. There is a cat avatar, like a Panther. If I have those on, }\\
        & & \textit{\footnotesize I’ll be called a furry, a lot." (T15, Echo VR)}\\ 
        
        \midrule
   
        \multirow{2}{*}{\small Physical} & \multirow{3}{*}{\small T8, T13, T17} 
        & \textit{\footnotesize "I had a deviated septum, where one part of the nasal passage is smaller than the other. So my voice sounded}\\
        \multirow{2}{*}{\small attributes} &  {}
        & \textit{\footnotesize really nasally, like, the stereotypical nerd voice. People would make fun of me because of that a lot." (T8, VRChat)} \\
        &   & 
        \textit{\footnotesize "Someone figured out I was mute and started going off on me making fun of my disability." (T17, VRChat)} \\
    \bottomrule	
\end{tabular}}
\end{table*}

%% file: sections/04-results-02.tex
\subsection{Users' Expectations for Safer VR}
\label{sec:user_expectations}

\noindent 
At the end of the interview, we asked participants about their expectations for safer VR experiences. Participants had insightful suggestions for improving safety controls and new ways of tackling harassment in VR.

\paragraphb{Live Moderators in VR.} Participants indicated a need for real-time assistance in various situations. T11 said: \textit{"There should be a guide, or an assistant to contact in case things go sideways"}. T17 wanted live moderators who were regular users capable of flagging those who violated the terms of service. T15 described moderators as "in-game security guards," while T12 compared them to the police:
\begin{myquote}
     "99\% of the time, most of us don't see a police person. But if we call one because we really need one, they come. If you could just push a button and have a person called to you, [they] can come and assess the situation." (T12)
\end{myquote}
T18 expressed outrage at the notion of social spaces without police officers and emphasized the need for social accountability. T15, a leader in a community for VR gamers (\textit{Virtual Reality Party League}), specified actionable ways to promote live moderation with the help of VR community leaders: 
\begin{myquote}
    "Have community leaders be involved in the main social aspects of games. Whether that is community leaders becoming mods or creating community members who want to step up and become mods. On the back end, have it built properly so they are properly trusted and educated on what they're supposed to do [with] their tools." (T15)
\end{myquote}
He also suggested offering perks to the moderators or having it as a paid position:\textit{"They might get a hierarchy rank in that development team. They're allowed to go to certain events. And they're really there for the idea of that community growth. Or it could be paid, you have to be on this headset hour by hour to cover this block and we'll pay you the sum amount"}.

\paragraphb{Tracking Users' Behaviour.} Participants suggested that VR apps and platforms track VR users' behavior. T1 felt that having access to every user's history was paramount, primarily to determine what action should be taken against them: \textit{"When they ascertain that this person has a history of harassment, [they] can give the person a warning first, and if it continues, just block the person completely, and make sure that the person will not have access to the platform again"}. T12's wanted users' behaviors to be tracked such that they had a certain "trust" level in social spaces inside an app:
\begin{myquote}
"It's good to have certain levels where you trust people or give them abilities or access to places, depending on whether they've earned that. If somebody has been reported for harassment, I'd be booting them right back to level one. They should have to earn back the privilege to be 'normal' again." (T12) 
\end{myquote}

T13 expressed that if a user was flagged for being toxic in one VR space, their \textit{"toxicity"} should \textit{"follow from one app to another"}. 
He felt that transferring repercussions across apps was essential for a fundamental shift in people's online behavior. He added that it had to be implemented at the platform (\textit{"Meta"}) level, with a coalition among popular VR apps:\textit{"The 10 most popular VR MMO games go, we're going to combine efforts and if you get three strikes combined in any of our games, you can't play any of our games anymore"}.

\paragraphb{Detect Distress in Users.} Some participants wanted their headsets to detect when they were feeling distressed. T11 felt that VR headsets should be sensitive enough to determine when a person needed help: \textit{"It should tell when a person needs help when he or she no longer feels comfortable in the game"}. T3 added that body movements such as rapid blinking of the eyes or sensors in the headset could be used as indicators of distress: \textit{"When I'm distressed, it gets this information and shuts down. If I am talking to you and I start blinking rapidly, it should know this person isn't okay, and slow down things"}.

\paragraphb{In-app Interventions.} Several participants indicated their preference for having in-app interventions such as disclaimers, warnings, or prompts to inform them before engaging in potentially toxic environments. T2, who encountered offending scenes on `YouTube VR,' wanted a disclaimer about the contents of a video. T11 suggested the inclusion of a prompt that would ask the user if they wanted to exit an app when harassment was detected:\textit{"If your headset can detect [harassment], there should be a prompt [asking] if you would want to leave the game you are in"}. T8 and T9 wanted features that would warn users before they entered social spaces where they might encounter harassers:
\begin{myquote}
 "If somebody is playing in a certain lobby and they'll know you, recognize you, and get angry at you again, I don't think the game should allow you to join it or tell you who is in the lobby before you join it. It should have a disclaimer that says you've tagged this person." (T8)
\end{myquote}
T9 added that users joining areas with problematic users (e.g., trolls) should get a notification warning them.

\label{sec:segregation_of_users}
\paragraphb{Segregation of Users.} Participants had concerns about the safety of kids and wanted age-based restrictions for VR usage: 
\begin{myquote}
    "VR shouldn't be for kids, honestly. Oculus is already requiring apps to remove 13-year-olds on it. I don't think they're going to stop using Oculus; they're just going to create a normal account and start communicating." (T5)
\end{myquote}
T18 added, \emph{"No children should be allowed to play with adults. Period. There needs to be age segregation, [because] there are adult predators transgressing boundaries of morality"}. T9 echoed the notion of segregation and suggested that the entire app be divided into age brackets: 
\begin{myquote}
    "There would be a bracket [with] young kids, another intermediate bracket, [with] teenagers. Adults, somebody will have to agree to some terms and conditions. If you're joining this section, know you may experience this and this. In other sections, all of that is banned." (T9)
\end{myquote}

\noindent T14 felt that paid users could be segregated from unpaid users to filter out children:\textit{"It would be cool if I could pay money to only play with people who paid money because it would get rid of a lot of screechers"}. T5, who had negative experiences with child users saying racial slurs and destroying virtual artifacts inside games, wanted adult-only lobbies. He also believed that checking IDs could be a way to enforce this:
\begin{myquote}
    "I'm not saying that kids shouldn't be able to enter, just that there should be a way that adults can just stick with each other, maybe like ID check." (T5)
\end{myquote}

\paragraphb{\underline{Takeaways:}} VR users believe that live moderators and age-based segregation would significantly reduce harassment in VR. They also recommend automatic detection of harassment situations and tracking users' toxicity histories across VR apps. Further, they want in-app interventions to inform them of harassers in the vicinity or in VR spaces they enter. 

%% file: sections/05-results-03.tex
\section{Developers' Perceptions on VR Safety}
\label{sec:dev_study_results}

\noindent Based on our findings from \textbf{Study-I}, we conducted \textbf{Study-II} (\S~\ref{sec:method}) where we interviewed (n = $\NumberOfDevInterviews$) VR developers. We asked them about the feasibility of the safety features desired by our users (\S~\ref{sec:dev_user_expectations}), challenges in designing and deploying safety controls (\S~\ref{sec:dev_challenges}), and ways to improve existing controls (\S~\ref{sec:dev_improving_safety}). With professional developers, VR-based researchers, and hobby developers represented by our participants  (Table~\ref{table:dev_demographics} in Appendix~\ref{sec:dev_demographics}), we believe our findings provide an extensive view of the challenges and limitations in this domain. 

\subsection{Feasibility of User's Expectations}
\label{sec:dev_user_expectations}

\noindent One portion of our interview protocol was about eliciting developers' perceptions of the features desired by the participants of \textbf{Study-I}. We started by asking the developers' perspective of what was needed to make VR safer and followed it by presenting the main user-desired features --- live moderators in VR, tracking users' behavior, detecting distress in users, in-app interventions, and segregation of users --- and asking about their feasibility. Since the segregation of users largely stemmed from users' desire to have an identity or age-based segregation, we asked developers about the feasibility of identity verification.

\paragraph{Live Moderators in VR -- Feasibility.} Developers largely agreed that having live moderators in every social space was infeasible due to the human resources required, the financial challenges, and the difficulty in scaling. D8 said: \textit{"That wouldn't be very scalable, especially for teams like mine where we only have three people developing"}. D7 argued that the economic model would not work, with D9 adding: \textit{"a small development studio, [with] six people and 100 servers couldn't fund that many people to do that. Even a big company who could do that, [would] have to fund thousands [of] people to listen in on every conversation that's said".}

To tackle these issues, D3 suggested leveraging the VR community to ease the burden on human moderation, echoing T15: \textit{"some platforms have community outreach, where respected members of the community can act as deputies of sorts"}. Drawing from his experiences in community management for his VR game, D7 added that live moderation was taxing and could be effective long-term only if moderators were part of a community: \textit{"They have to get something out of the experience in the first place, or it's just something that burns people out"}. D3 advocated using a small, well-trained team to ensure consistent moderation practices. Multiple developers echoed that moderation at scale could be done using AI-based abuse detection but also recognized the inherent technical and privacy challenges:
\begin{myquote}
"Is it practical? How will people react to it? These are the questions that we need to be asking because we're already doing voice, but what about content and avatars? When are we going to have models that look at a 3D mesh and identify if there's any perfect content on it?" (D3)
\end{myquote}

\paragraphb{Tracking Users’ Behaviour -- Feasibility.} Developers argued that tracking users' behavior was feasible from a technical standpoint; however, they did not favor its deployment. D1 pointed out that if a user violated the terms and conditions of a VR community once, it may not be fair to exclude them from other apps, especially if they corrected their behavior. D9 added: \textit{"If you connect every app ever and you're toxic on one app and it now spreads to everything, obviously people wouldn't want that. You could also have silent toxic people that just lower your rating"}. D3 further highlighted the importance of ensuring that users did not get a negative impression of a user before interacting with them: \textit{"I would not put [the rating] in something which is always stuck to your head. But if I made the conscious decision to require more information about [the user], it's present"}.

Developers highlighted potential challenges in implementing this technique. D8 believed that the tracking could happen only at the platform level (such as Meta, Steam, etc.) due to the existence of several VR stores, each using different accounts. D6 added that users must be incentivized to use the same profile on all apps/platforms. However, they also found this solution to be privacy-infringing: 
\begin{myquote}
    "As a user, I don't like to be pervasively tracked by corporations. Whether or not they're providing me tangibly usable tools, they're not trustworthy entities." (D7)
\end{myquote}
D2 argued that, while the data could be anonymized if it was used for training an AI model to detect harassment, real-time tracking involved legal challenges. D3 wanted VR companies to be transparent about data collection practices to users: \textit{"Transparency on the company's part is necessary to reaffirm faith in users"}. Despite their thoughts on how this solution compromised users' privacy, developers also perceived the value of tracking users:
\begin{myquote}
    "It's good to have that unified information. As a developer, I would love to have a jump list of problematic users and an ability to just, at the very least, pre-sort them into [a similar] category of servers." (D7)
\end{myquote}

\paragraphb{Detect Distress in Users -- Feasibility.} Our participants agreed that the automatic detection of distress in users to turn off the headset or quit a VR app seemed a good solution but raised concerns about the availability of sufficient input for accurate detection:
\begin{myquote}
    "We'll have to be able to break down the feasibility in terms of being able to understand when any stress is happening, how to interpret it, in order to implement it." (D2)
\end{myquote}
\noindent D4 believed there could be many false positives, where emotions such as excitement may be wrongly detected as distress. D6 added that false positives would result in a \textit{"terrible user experience"}. D9 suggested that obtaining feedback from the user upon detecting distress might achieve a good trade-off between model improvement and user experience.

\paragraphb{In-app Interventions -- Feasibility.} The developer participants agreed that having in-app interventions was \emph{feasible} solution. D1 suggested the use of voice-activated commands for interacting with pop-up notifications and encouraged designs with minimal visual interaction: 
\begin{myquote}
    "The less you have to disturb the user in terms of having to do manual input, that's good. You can [use] pop-ups and [they] just [have] to click, that's useful. It can even be voice-activated commands [without the] user having to interact too much with hand controls." (D2)
\end{myquote}

\paragraphb{Identity Verification -- Feasibility.} Since our user participants wanted to segregate users based on age or identity verification, we asked the developers about their perspectives on implementing identity verification. They felt that users would not be willing to share official identification as it was invasive, compromised anonymity, and posed risks in cases of data breach while failing to address the problem:
\begin{myquote}
    "Someone who's someone is immaterial to whether or not they are a problem. Having their government ID is just putting more power in the hands of tech platforms." (D7) 
\end{myquote}

\noindent D1 and D2 suggested the integration of social media profiles to VR accounts; however, such initiatives have received backlash in the past, as described by D9: \textit{"[Meta] were trying to connect everything from Facebook, Instagram, Oculus, and they got in trouble for doing that"}. D2 and D5 recommended using hardware identifiers or IP addresses to detect a user. D2 suggested third-party verification but also highlighted challenges in complying with different technology laws across countries. D3 shared that some VR clubs required real-life identification to partake in the community:
\begin{myquote}
    "Some communities have third-party services to check if the person's real. And that can vary. You take a selfie with your ID, let your personal information out. Or it can be a Discord bot that takes a picture of your driver's license, talks to a background checking service." (D3)
\end{myquote}

\noindent D5 suggested computer vision techniques for face detection to predict a user's age, while D6 recommended using physiological attributes such as movement and posture for inference.

\paragraphb{\underline{Takeaways:}} Sustainable live moderation in VR communities may rely on moderators embedded in the community and augmented with AI-based abuse detection. While tracking users would enable VR developers to create a block list of problematic users, it may be limited to a single platform (e.g., Meta, Steam, etc.) and pose legal risks to VR companies. Detecting distress in users requires a thorough evaluation of whether VR systems can access sufficient input for accurate detection. Developers advocate third-party verification for non-invasive user identification and propose in-app interventions.

\subsection{Challenges in Designing Safety Controls}
\label{sec:dev_challenges}

\noindent Several questions in our interview were about extracting the challenges in the development of safety controls. We asked developers whether and when safety control design appeared in the pipeline of the apps they developed and what, according to them, was hindering the development of effective safety controls in VR apps. We also asked them if they believed an industry standard of VR safety controls could be created and what technical, social, and economic barriers they perceived.

\paragraphb{Safety Not A Priority.} A majority of our developer participants believed that VR platforms and app developers do not currently prioritize safety. D3 explained that moderation in VR tended to be reactive rather than proactive, as the development efforts would be focused on appealing to the investors and clients. D4 and D8 felt that VR developers usually had small teams and did not have the bandwidth to develop features for safety. D9 added that developers would not implement safety features until their users reported issues:
\begin{myquote}
    "Companies are very money-first, fix later. [Maybe] this is why it's getting pushed off, and not many people are talking about it or fixing it." (D9)
\end{myquote}

\noindent Several participants agreed that there was a lack of financial motive to prioritize safety. D7 highlighted the need for tech companies to justify developmental efforts, especially in light of recent economic conditions with mass layoffs: \textit{"the primary challenge is justifying all that development effort when that could be put towards new user acquisition or directly monetizing engagement"}. D7 added that big companies did not stand to lose out monetarily due to the users that stopped using apps due to harassment:
\begin{myquote}
    "You can always try to acquire new users faster than you are losing older users. If [companies] can compensate for X women who are leaving, and it's just awful being a woman online, period, but have a referral program that gets more teenage boys to recommend the platform, and they get a \$3 kickback and that makes their numbers go [up], then [they] don't have a problem." (D7)
\end{myquote}

\noindent Participants also believed that well-implemented safety controls may limit interactions among users, impact overall app engagement, and disincentivize companies to focus on them:
\begin{myquote}
    "Platforms wouldn't wanna do that because engagement is easiest to generate via conflict. If you actually allow people to peaceably exist separate from each other, there are repeat logins that people don't end up doing." (D7)
\end{myquote}

\paragraphb{Lack of Guidelines/Standards.} Developers argued that they did not have any guidelines for ensuring safety in VR apps and stressed the importance of creating awareness among developers to design for safety. For instance, D5 shared:
\begin{myquote}
    "We don't have a list to go through [that] says you should have [these] models in your app for safety. I'm not sure if it's already there. But clearly, it's not in my mind. It would be nice to have  clear regulations to follow." (D5)
\end{myquote}

\noindent When asked about the feasibility of creating an industry-wide standard for VR safety controls, most of our participants foresaw challenges in bringing together the major stakeholders in the ecosystem. D2 felt that effective enforcement of a standard required coordination among all the big companies, while D7 illustrated how companies tended not to cooperate:
\begin{myquote}
    "Apple goes out of their way to not use any of the same words to describe things. They just announced a VR headset without using the words virtual reality once. It's antithetical for them that there's some collective best way to handle user moderation. It had to be forced by the EU to put a charger on their phone. That tells how much these corporations want to cooperate on anything." (D7)
\end{myquote}

\noindent D2 illustrated the technical challenges in implementing uniform data collection practices across platforms: \textit{"To implement such a standard, the way data is gathered in application A should be similar to application B. You have to develop [a] middle framework capable of [collecting] this data and parsing it"}. D6 felt that enforcing entities to adhere to standards was hard. D5 believed non-profit organizations could facilitate companies working together.

\paragraphb{Practical Challenges.} Developers identified several technical and logistical challenges in the design and deployment of effective safety mechanisms for VR. D2 felt, \textit{"it's complicated to prove that someone is harassing you because it's through the internet, you have the VPN, can hide your IP. It's tricky to prove that someone is harassing you"}. D7 stressed the importance of recruiting good moderators for his VR game:
\begin{myquote}
   "You need effective moderators. People who are aware of the rules, who see eye to eye to you about what they mean, why they exist, what your goals for the community are, who you wish to include and exclude in that community, and willing to put in the hours being present. They shouldn't just feel like silent overlords because that can build resentment patterns between user bases." (D7) 
\end{myquote}

\noindent An inherent challenge about safety controls that developers called out was that they offloaded a lot of responsibility to the user, as users needed to remember the right controls and use them at the time of need. This is further exacerbated by complications in usability testing, specifically in replicating stressful situations to test how usable the tools are: 
\begin{myquote}
    "The biggest barrier is actually simulating those experiences to see the tools that are actually working." (D8)
\end{myquote}

\noindent D7 also emphasized the complicated interface development work involved in synthetically recreating reality for VR apps. 

\paragraphb{Compromise Privacy.} Most of our participants expressed concerns about using methods that infringed privacy for the sake of safety and questioned the necessity of invasive practices for moderation. D3 said, \textit{"most people are turned off by the idea of [something] constantly monitoring them, possibly storing their conversations on a database somewhere for an extended period of time"}. While discussing ways to collect evidence for reporting, many developers suggested using a video buffer that stored the recording from a specific time period (e.g., the last 10 seconds); however, all of them agreed that it was privacy-violating:
\begin{myquote}
    "Say the platform has the recording of everything in the last 24 hours. The user can always go back to see what happened around [them]. So everything will be recorded, but I don't know how that will conflict with privacy." (D5) 
\end{myquote}

\paragraphb{\underline{Takeaways:}} VR companies tend not to prioritize safety due to a lack of financial incentives and high development costs. Further, VR developers lack awareness about safety risks in VR and may not have legal or technical guidelines for safety design. Developers also highlight challenges in simulating VR safety risks for user testing of the safety controls and raise questions about balancing privacy with safety.

\subsection{Improving Safety in VR}
\label{sec:dev_improving_safety}

\noindent We asked developers about the similarities and differences in dealing with harassment in VR when compared to traditional forms of social media. We then asked them what they perceived to be lacking in safety controls and how they may be improved. We also presented the usability challenges from \textbf{Study-I} and asked developers how they would solve them.

\paragraphb{Abuse Detection.} Most of our developer participants agreed on the challenges involved in moderation at scale and recommended using several abuse detection methods as the first step in the moderation life cycle. D3 believed, \textit{"there's certainly the chore that is the actual implementation, once those systems are in place, they're largely autonomous. I don't need to touch it."}, and explained how word analyzers may be used to identify verbal abuse and notify users:
\begin{myquote}
    "Using semantic analyzer, we have been able to identify the intent of what someone is saying. If we identify that as harmful or having foul language, we can give the user a reminder, like, hey, you should be watching your language. If it persists, we can take moderation action." (D3)
\end{myquote}

\noindent D2 added that a common list of bad words may be detected and used to notify relevant users: \textit{"We're thinking of being able to remove some words from the usage of the users"}. For detecting other forms of abuse in a non-invasive manner, D6, a VR-based researcher, proposed the use of logged events:
\begin{myquote}
    "I would keep track of user events that can occur, positions of their avatars, the proximity of different avatars. If I have that record, even without video evidence, I would know if they were in proximity, if some user events occurred. If I have the username reported, I can see if they were in those proximity and had the opportunity to interact." (D6)
\end{myquote}

\paragraphb{Effective Grouping of Users.} Drawing on his experience running a VR game, D7 shared that effective community management with like-minded users grouped together prevented conflicts, reinforcing users' preferences from \textbf{Study-I}:
\begin{myquote}
    "One of the bedrocks of sustainable community management is about establishing a set of norms and expectations and creating an environment that funnels those correctly so that you are effectively grouping users together who share expectations. And if they're sharing expectations, you don't get people in conflict [like when] one person thought they were here for X and another thought they were there for Y, and they're now fighting over that." (D7)
\end{myquote}

\noindent Developers further elaborated on interaction filtering and social graph pruning to group users. D7 said: \textit{"I'm in party mode. I hit a button, and it makes my social protocols in terms of who can talk to me, and this becomes broad momentarily because I trust the people running this party. Then I switch to a different virtual space, and I hit a button that changes who can DM me for a private chat, who appears on my screen"}. He added that all connections of specific users should be prevented from ever interacting with another user to effectively cut off problematic social graphs. 

\paragraphb{Solving Usability Challenges.} To solve some of the usability challenges that our participants faced in \textbf{Study-I} (pointing at a user in virtual space to block them, choosing a user from a long list of usernames, typing on a keyboard in VR), developers proposed several potential solutions, and generally advocated for designing intuitive controls that felt natural to use. D3 believed that blocking tools needed more granularity, providing the capability to block different aspects of a user, such as \textit{"their voice, avatar, ability to scale, and other atomic elements that users can have fine-tune control over"}. D7 added that users should be able to block other users for different periods of time: \textit{"you should be able to shut someone up for 10 minutes if they're just being annoying"}. 

To effectively identify users for taking action, D4 recommended designing ways to pause the scene and recalled how Meta Horizon Worlds implemented this: \textit{"you just hold up your arm and press a button. It immediately freezes everybody, but you still see other people in the room. If someone is harassing you, you can pause everything, report [or] block them"}. D7 added that the list of users can be accompanied by a snapshot of their avatars:
\begin{myquote}
    "If that list is difficult to deal with, why is that list not sorted by distance to you physically? Or grouped into sets of bands? You're trying to identify someone like oh it's a guy with a hat and a coat. If you think about the way you would search a physical space in that context and mark people off, effectively modeling that cognitive process as a UI tool is a way to do that." (D7)
\end{myquote}

\noindent D8 also recommended implementing voice-based memos for actions such as reporting in order to avoid typing in VR. 

\paragraphb{Creating Awareness among Developers.} Three out of nine developers who took part in \textbf{Study-II} argued that more awareness was needed among VR developers about harassment issues in VR. D9 expressed that independent developers needed to be made aware of safety issues and provided with the relevant resources for them to adopt a safety-oriented design:
\begin{myquote}
    "When I was developing, the last thing I was going to add was safety features. So bringing more awareness to the little guys, the small developers, giving them resources that allow them to integrate these things quickly \dots" (D9)
\end{myquote}
D8, who was part of a three-member development team, argued, \textit{"if it was a big enough issue, even though we're a small team if it was brought to our attention, probably have a sprint to focus on that, at least one of us"}.

\paragraphb{Open-sourced Safety Libraries.} Several participants shared that open-sourced libraries available through popular game engines that allowed developers to integrate standard safety controls into their apps would be valuable. D9 said: \textit{"if you had a library you could pull from, easy implementation. Especially if it just works out of the box"}. He further added that if it were open-source, when somebody found a way to break it, the community would be able to fix it. D2 added:
\begin{myquote}
    "It should be a plug-and-play tool that could be brought to the game engines, and you just drag and drop." (D2)
\end{myquote}

\noindent D8 pointed out the importance of having well-established requirements in app stores for targeting efforts towards safety: \textit{"If some stores have requirements where your app needs [certain] features to be admitted, that would force all the devs to have at least something to protect its users"}. D1 and D7 stressed that seminars and workshops may be organized for developers to brainstorm on creating such tools. D5 also suggested creating standardized tutorials for VR users, informing them how to protect themselves. 

\paragraphb{\underline{Takeaways:}} Using word filters and semantic analyzers may aid in abuse detection. Developers recommend implementing extensive controls to limit or allow user interaction in various settings. Granular controls allow users to restrict interaction temporally and spatially address certain limitations of existing safety measures. Seminars and workshops can be conducted to inform developers about VR safety and foster innovation in safety design. Additionally, integrating open-sourced safety-focused libraries with popular game engines can streamline the development of safety controls.

%% file: sections/06-discussion.tex
\section{Discussion}
\label{sec:discussion}

\noindent In this paper, we outlined our findings about VR safety features based on a multi-perspective study with \NumberOfUserInterviews targets of VR-based harassment and \NumberOfDevInterviews VR developers. In this section, we discuss the broader findings from our study and offer recommendations to help VR platform owners, app developers, and policymakers enhance VR user safety.

\paragraphb{What's New With VR-based Harassment?} 
Much like various other manifestations of online harassment, VR-based harassment exposes users to harmful content, including bullying, threats of violence, and sexual harassment~\cite{thomas2021sok}. While certain forms of harassment may be universally applicable across different mediums, the distinctive features of VR applications give rise to unique manifestations of harassment. For instance, incidents such as virtual sexual harassment, where the harasser positions their avatar's "crotch" on the target's face, and virtual violence, where a harasser can manipulate the target's avatar to cause disorientation, illustrate the specific challenges posed by VR, as outlined in Table~\ref{tab:taxonomy_of_harassment}. 

VR parallels other forms of online interaction, such as gaming, so many solutions that apply to the latter (profanity filters, reporting systems, behavioral analytics, etc.) apply to the former. However, implementing the same solutions effectively in VR requires significant redesign to accommodate the three-dimensional and immersive nature of interaction that leads to more traumatizing forms of harassment (e.g., virtual sexual harassment). Further, safety controls are not equally effective across VR contexts, particularly while removing harassers from gaming and streaming settings, highlighting the uniqueness of safety design for VR. Proposed solutions have included robust reporting mechanisms~\cite{freeman2022disturbing}, non-player characters as safety companions~\cite{zheng2023safetyrisks}, and consent-based designs~\cite{zytko2023dating, schulenberg2023we} for boundary settings. We add to this by identifying ways to improve safety in VR --- keeping the bad actors in check while minimizing the load on users and enabling developers to implement these solutions. 

\paragraphb{Minimizing Load On Users.} With the burden for staying safe online currently falling on users~\cite{wei2023there}, the fact that many participants from \textbf{Study-I} were unaware of VR safety controls calls for improving user awareness about safety controls. Apart from making the information available in-app as tutorials, nudges to users, and displaying cues in VR in the style of "Madison Avenue" advertising \cite{arzaghi2008networking}, we also recommend that VR app/platform owners inform users through official websites, app stores, and social media campaigns. However, as our findings highlight, understanding the usage of safety controls is a time investment that VR users must be willing to make. A good design of safety controls should make this learning process seamless and consider particular contexts in which users would employ them, such as gaming or streaming settings. Standardizing the baseline safety controls across the VR ecosystem may help, as participants report that safety controls get easier to use with familiarity.

\paragraphb{Stopping Bad Actors.} Stopping bad actors requires automated or human-driven identification of bad behavior and enforcement of sufficient, non-biased punitive action. The action may range from warnings and penalties to bans that prevent harassers from returning to the platform~\cite{blackwell2019harassment}. While several ML-based techniques have been developed to detect obscene imagery~\cite{tahir2019bringing} and abusive language~\cite{chandrasekharan2017bag}, there is a need to direct efforts towards detecting a variety of toxic content from a 3D environment.

Since user reports play a major role in VR moderation, designing usable reporting mechanisms with multi-modal availability is critical. Again, standardizing the reporting features across VR would reduce the burden on users. Effective moderation requires VR apps/platforms to have diverse, well-trained moderators in the community. AI-based moderation techniques~\cite{schulenberg2023towards} may be considered for prompt processing and providing timely feedback to the reporter. Models of norm enforcement such as \textit{responsive regulation} have been identified in the literature~\cite{blackwell2019harassment}, where the penalties are proportional to the offense's severity and the perpetrators' intent.

Identifying cases of false reporting and providing constructive feedback to the ones reported for corrective behavior should be an integral component of moderation. Prior work has proposed the inclusion of a "limbo" space where a reported user could be taken to while the moderator provides them feedback about why they were reported~\cite{sabri2023challenges}. Non-invasive forms of identity verification need to be implemented to prevent toxic users from returning to the platform in case of a permanent ban. 

\paragraphb{Enabling Developers.} A key part of solving the safety problem is enabling VR developers. The design of safety controls requires innovation in usability as they need to be usable in three-dimensional space, in contrast to their counterparts in other online social platforms. As more and more inexperienced developers enter the burgeoning VR industry, awareness is critical to ensure a safety-oriented design of VR apps. Although the VR ecosystem consists of a diverse population of independent developers, gaming companies, and corporations, likely with conflicting priorities~\cite{blackwell2019harassment}, technical and legal guidelines mandating standards for safety and data collection practices would ensure cross-platform compliance and reduce the burden on developers. VR-specific regulations for appropriate codes of conduct, tutorials, moderation systems, and the required set of safety features in VR platforms/apps would simplify the design process for developers. Open-source development of VR safety-oriented libraries would add immense value, particularly if integrated into game engines.

\paragraphb{Considering Multiple Perspectives.} Although VR users and developers had shared opinions in our study, such as using community members for effective moderation, developers expressed more concern about users' privacy compared to users while thinking about safety design for VR. Although users recommended behavioral tracking across VR apps to ensure safety, by implementing these pervasive tracking mechanisms, VR app companies may be subject to legal scrutiny and lose their users' goodwill in cases of false reporting. From \textbf{Study-II}, developers had conflicting views on when safety controls needed to be included in the development pipeline for a VR app: some wanted to design them at the start to be in tune with the affordances of the app, while others wanted to focus on the core functionality first. Although developers recommended safety-focused research, they also highlighted the challenges of simulating harassment scenarios for user testing. It is imperative to reflect on these contrasting perspectives to implement solutions that achieve a good balance of safety and privacy with ease of development. 

%% file: sections/07-conclusion.tex
\section{Conclusion}
\label{sec:conc}

\noindent Since VR serves as an emulation of the real world, it presents many of the challenges in society, perhaps creating a notion that harassment in VR is a societal problem. However, VR-based harassment is yet another instance where technology facilitates malicious social actors to thrive, similar to social engineering attacks such as phishing or robocalls. Thus, similar to developing solutions for phishing or robocalls, we must engineer robust solutions to address VR-based safety risks. Techniques such as AI-based abuse detection, semi-automated moderation, and identity verification raise questions about users' privacy; therefore, it is vital to identify practical, privacy-preserving solutions that serve VR platforms, developers, and users. To allow for anonymous VR interactions while creating safe spaces, VR apps may create tiered experiences, with some spaces having identity verification and others not. To prevent vulnerable populations such as children from partaking in potentially dangerous VR spaces, users' gait, height, and facial features may be used to run a model on-device to infer their age. Problematic and disruptive users may be removed from other users' views by pruning their social graphs. The onus is on major players in the VR ecosystem to come together to solve the problem of harassment in virtual environments. 

%% file: sections/08-ack.tex
\section*{Acknowledgments}

\noindent We are deeply grateful to the VR users and developers who participated in our studies, as well as Dr. Vanessa Volpe, Department of Psychology, North Carolina State University, who helped us refine Study-I. We extend our thanks to Elizabeth Lin, Greg Tystahl, Sathvik Prasad, and Zahra Shiraz for their feedback on the interview protocols. We also thank our anonymous reviewers and shepherd for their constructive and insightful feedback, which helped improve this paper significantly. This material is based upon work supported in parts by a Meta Research gift award. Any opinions, findings, conclusions, or recommendations expressed in this material are those of the authors and do not necessarily reflect the views of Meta Inc. 

%% file: main-arxiv.bbl

%% file: sections/09-appendix.tex
\onecolumn

\appendix 

\section*{Appendix}

\section{Participant demographics}

\subsection{Targets of VR-based harassment}
\label{sec:user_demographics}
\input{tables/user_demographics}

\vspace{8pt}

\subsection{VR developers}
\label{sec:dev_demographics}
\input{tables/dev_demographics}

%% file: tables/user_demographics.tex
\begin{table*}[!h]
	\centering
 \caption{Demographic information of targets of VR-based harassment (self-reported).} 
 \resizebox{1.0\textwidth}{!}{
		\begin{tabular}{cccccccccc} 
			\toprule 
			 
   \multirow{2}{*}{ID} & \multirow{2}{*}{Age} & \multirow{2}{*}{Gender} & \multirow{2}{*}{Sexual Orientation} & \multirow{2}{*}{Race} & Usage & Usage & \multicolumn{3}{c}{VR apps with harassment experience}\\
   \cmidrule(r){8-10}
   &  &  &  &  & (years) & (hrs/week) & Social VR & Gaming VR & Streaming VR\\ 
		
   \midrule
  {T1} &
  {18-24} &
  {Non-binary} &
  {Other} &
  {Black} &
  {6} &
  {4-20} &
  {} &
  {Asgard's Wrath} &
  {}
 \\ \midrule
{T2} &
  {25-34} &
  {Male} &
  {Heterosexual} &
  {White} &
  {5} &
  {4-20} &
  {} &
  {} &
  {YouTube VR}
 \\ \midrule
{T3} &
  {25-34} &
  {Female} &
  {Heterosexual} &
  {Black} &
  {4} &
  {20-40} &
  {Second Life} &
  {} &
  {} 
\\ \midrule
{T4} &
  {25-34} &
  {Female} &
  {Heterosexual} &
  {Black} &
  {3} &
  {} &
  {} &
  {Beat Saber} &
  {BigScreen}
  \\ \midrule
{T5} &
  {18-24} &
  {Male} &
  {Heterosexual} &
  {Black} &
  {3} &
  {4-20} &
  {VR Chat} &
  {Beat Saber} &
  {}
\\ \midrule
{T6} &
  {25-34} &
  {Male} &
  {Heterosexual} &
  {White} &
  {<2} &
  {4-20} &
  {} &
  {Star Wars: Squadron} &
  {YouTube VR}
\\ \midrule
{T7} &
  {45-54} &
  {Female} &
  {Other} &
  {White} &
  {3} &
  {4-20} &
  {} &
  {Orbus, Zenith} &
  {}
\\ \midrule
{T8} &
  {18-24} &
  {Male} &
  {Bisexual} &
  {White} &
  {4} &
  {4-20} &
  {VR Chat} &
  {Pavlov VR} &
  {}
\\ \midrule
{T9} &
  {18-24} &
  {Male} &
  {Heterosexual} &
  {Black} &
  {4} &
  {4-20} &
  {Sinespace} &
  {Archangel} &
  {}
\\ \midrule
{T10} &
  {25-34} &
  {Female} &
  {Heterosexual} &
  {Black} &
  {<1} &
  {1-4} &
  {Second Life} &
  {} &
  {}
  \\ \midrule
{T11} &
  {25-34} &
  {Male} &
  {Heterosexual} &
  {Black} &
  {<1} &
  {1-4} &
  {VR Chat} &
  {House of Terror} &
  {YouTube VR} 
  \\ \midrule
{T12} &
  {45-54} &
  {Female} &
  {Heterosexual} &
  {White} &
  {6} &
  {20-40} &
  {} &
  {Echo VR} &
  {} 
  \\ \midrule
{T13} &
  {35-44} &
  {Male} &
  {Heterosexual} &
  {White} &
  {2} &
  {4-20} &
  {VR Chat} &
  {Echo VR} &
  {} 
  \\ \midrule
{T14} &
  {35-44} &
  {Male} &
  {Other} &
  {American Indian} &
  {7} &
  {4-20} &
  {VR Chat} &
  {Echo VR} &
  {Big Screen}
  \\ \midrule
{T15} &
  {25-34} &
  {Male} &
  {Heterosexual} &
  {White} &
  {4} &
  {20-40} &
  {VR Chat} &
  {Echo VR, Gorilla Tag} &
  {} 
  \\ \midrule
  \multirow{2}{*}{T16} &
  \multirow{2}{*}{18-24} &
  \multirow{2}{*}{Male} &
  \multirow{2}{*}{Heterosexual} &
  \multirow{2}{*}{White} &
  \multirow{2}{*}{1} &
  \multirow{2}{*}{1-4} &
  \multirow{2}{*}{RecRoom} &
  {Echo VR, Pavlov VR,} &
  {} 
  \\
  {} &
  {} &
  {} &
  {} &
  {} &
  {} &
  {} &
  {} &
  {Walkabout Mini-Golf} &
  {} 
  \\ \midrule
{T17} &
  {18-24} &
  {Female} &
  {Heterosexual} &
  {American Indian} &
  {<1} &
  {4-20} &
  {VR Chat} &
  {Echo VR} &
  {} 
  \\ \midrule
{T18} &
  {35-44} &
  {Male} &
  {Homosexual} &
  {Other} &
  {4} &
  {20-40} &
  {} &
  {Echo VR} &
  {}
  \\ \bottomrule
    \end{tabular}}
\label{table:demographics}
\end{table*}

%% file: tables/dev_demographics.tex
\begin{table*}[!h]
	\centering
 \caption{Demographic information of VR developers (self-reported).} 
 \resizebox{1.0\textwidth}{!}{
		\begin{tabular}{cccccccccccc} 
			\toprule 
			 
   \multirow{2}{*}{ID} & \multirow{2}{*}{Age} & \multirow{2}{*}{Gender} & \multirow{2}{*}{VR dev} & \multirow{2}{*}{Education} & \multirow{2}{*}{Tools} & \multicolumn{6}{c}{Roles performed}\\
   \cmidrule(r){7-12}
   &  &  & experience & (degree) & used  & UI/UX & XR Gameplay & Software & Researcher & AR/VR & Other\\ 
   &  &  &  &  &  & Designer & \& Tools & Developer & & Maintenance & Roles\\
   &  &  &  &. &  &  & Engineer & & & \& Support & \\
   \midrule
  {D1} &
  {$25-34$} &
  {Male} &
  {$>=4$ years} &
  {Bachelor's} &
  {Unity} &
  {$\checkmark$} &
  {} &
  {$\checkmark$} &
  {} &
  {$\checkmark$} &
  {}
 \\ \midrule
  {D2} &
  {$25-34$} &
  {Male} &
  {$2-3$ years} &
  {Doctoral} &
  {Unity} &
  {} &
  {$\checkmark$} &
  {$\checkmark$} &
  {$\checkmark$} &
  {} &
  {Graphics}
       \\
   {} & {} & {} & {} & {} & {} & {} & {} & {} & {} & {} & {Engineer}
 \\ \midrule
  {D3} &
  {$18-24$} &
  {Male} &
  {$1-2$ years} &
  {Bachelor's} &
  {Unity} &
  {$\checkmark$} &
  {$\checkmark$} &
  {$\checkmark$} &
  {$\checkmark$} &
  {$\checkmark$} &
  {}
 \\ \midrule
  {D4} &
  {$35-44$} &
  {Male} &
  {$>=4$ years} &
  {Bachelor's} &
  {Unreal Engine} &
  {} &
  {$\checkmark$} &
  {} &
  {} &
  {$\checkmark$} &
  {}
   \\ \midrule
  {D5} &
  {$25-34$} &
  {Female} &
  {$3-4$ years} &
  {Master's} &
  {Unity} &
  {$\checkmark$} &
  {} &
  {} &
  {$\checkmark$} &
  {} &
  {}
  \\ \midrule
  {D6} &
  {$25-34$} &
  {Male} &
  {$>=4$ years} &
  {Master's} &
  {Unreal Engine} &
  {} &
  {} &
  {} &
  {$\checkmark$} &
  {} &
  {}
 \\ \midrule
   {D7} &
   {$35-44$} &
   {Male} &
   {$>=4$ years} &
   {Master's} &
   {Unity, Maya} &
   {$\checkmark$} &
   {$\checkmark$} &
   {$\checkmark$} &
   {} &
   {} &
   {Product,}
   \\
   {} & {} & {} & {} & {} & {Creative cloud,} & {} & {} & {} & {} & {} & {Manager,}
      \\
   {} & {} & {} & {} & {} & {Substance suite} & {} & {} & {} & {} & {} & {Marketing}
  \\ \midrule
  {D8} &
  {$18-24$} &
  {Male} &
  {$1-2$ years} &
  {Vocational} &
  {Unity,} &
  {$\checkmark$} &
  {$\checkmark$} &
  {$\checkmark$} &
  {} &
  {} &
  {}
     \\
   {} & {} & {} & {} & {training} & {Blender} & {} & {} & {} & {} & {} & {}
  \\ \midrule
  {D9} &
  {$18-24$} &
  {Male} &
  {$1-2$ years} &
  {Bachelor's} &
  {Unity} &
  {$\checkmark$} &
  {$\checkmark$} &
  {$\checkmark$} &
  {} &
  {} &
  {}
  \\ \bottomrule
    \end{tabular}}
\label{table:dev_demographics}
\end{table*}